# Archimedean Proof of the Physical Impossibility of Earth Mantle Convection

by

J. Marvin Herndon
Transdyne Corporation
San Diego, CA 92131 USA
mherndon@san.rr.com

**Abstract:** Eight decades ago, Arthur Holmes introduced the idea of mantle convection as a mechanism for continental drift. Five decades ago, continental drift was modified to become plate tectonics theory, which included mantle convection as an absolutely critical component. Using the submarine design and operation concept of "neutral buoyancy", which follows from Archimedes' discoveries, the concept of mantle convection is proven to be incorrect, concomitantly refuting plate tectonics, refuting all mantle convection models, and refuting all models that depend upon mantle convection.

## 1 Introduction

Discovering the true nature of continental displacement, its underlying mechanism, and its energy source are among the most fundamental geo-science challenges. The seeming continuity of geological structures and fossil life-forms on either side of the Atlantic Ocean and the apparent "fit" of their opposing coastlines led Antonio Snider-Pellegrini to propose in 1858, as shown in Fig. 1, that the Americas were at one time connected to Europe and Africa and subsequently separated, opening the Atlantic Ocean (Snider-Pellegrini, 1858).

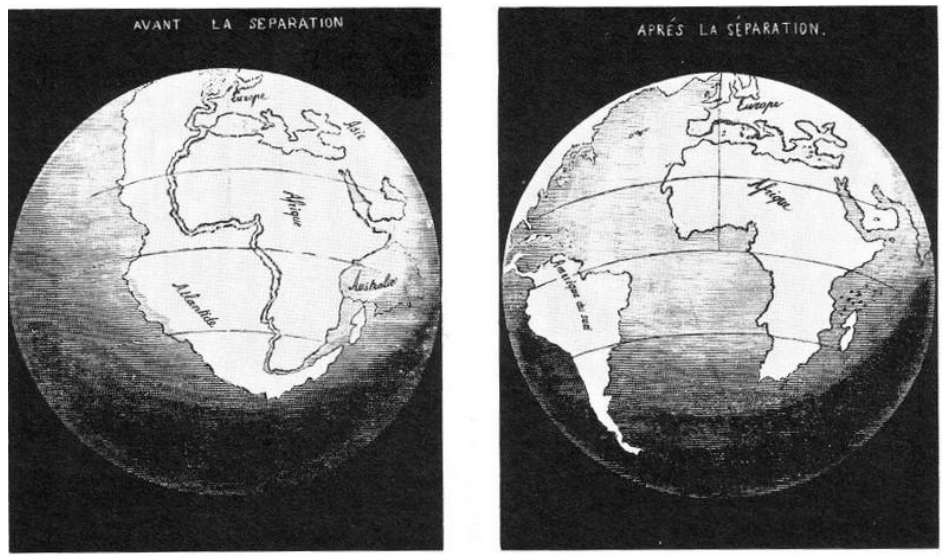

**Fig. 1** The opening of the Atlantic Ocean, reproduced from (Snider-Pellegrini, 1858).

Half a century later, Alfred Wegener promulgated a similar concept, with more detailed justification, that became known as "continental drift" (Wegener, 1912). Another half century later, continental drift theory was modified to become plate tectonics theory (Dietz, 1961;Hess, 1962;Le Pichon, 1968;Vine and Matthews, 1963).

Any theory of continental displacement requires a physically realistic mechanism and an adequate energy source. In 1921, Bull suggested the idea of mantle convection being involved in mountain formation (Bull, 1921). In 1931, Holmes elaborated upon the concept of mantle convection and suggested it as a mechanism for continental drift, publishing the illustration reproduced as Fig. 2 (Holmes, 1931). Mantle convection was later adopted as an inextricable part of plate tectonics theory, as illustrated by the U. S. Geological Survey diagram reproduced as Fig. 3.

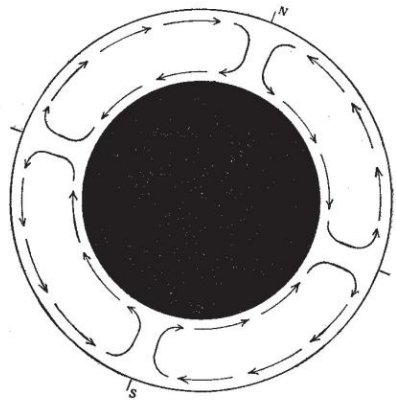

**Fig. 2** Schematic representation of mantle convection, from (Holmes, 1931). Reproduced with permission of the Geological Society of Glasgow.

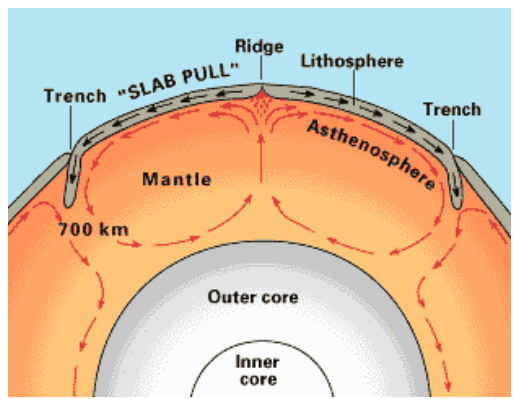

**Fig. 3** U. S. Geological Survey schematic representation of mantle convection associated with plate tectonics theory.

For eight decades, mantle convection has been taken for granted, its existence assumed without proof and justified by (flawed) numerical calculations. Herndon (2009, 2010a, b) first discovered the physical impossibility of mantle convection, revealed the widespread mis-application of the Rayleigh Number to the Earth's mantle, and pointed to flaws underlying mantle convection models.

Modeling mantle convection typically involves complex computer-mathematical calculations based upon approximations applied to the Eulerian equations of fluid dynamics and generally begins with the (false) assumption that mantle convection occurs in nature. Whereas sophisticated calculations may appear elegant and impressive, their use can be misleading, especially when parameterization techniques are applied. Rather than moving toward increased complexity and abstraction, the author attempts to move toward increased simplification and ease of understanding, reducing a problem to its fundamental elements, intimately connected to the properties and behavior of matter. In the following section, the author proves in a new, readily understandable, yet mathematically precise way, that convection in the Earth's mantle is physically impossible. The author then points to popular research topics that are no longer valid as a consequence of their being critically dependent upon physically-impossible mantle convection, and refers to a different development of geodynamics which is independent of mantle convection.

## 2 Physical Impossibility of Mantle Convection

The lava lamp, invented by Smith (1968), affords an easy-to-understand demonstration of convection at the Earth's surface. Heat warms a blob of wax at the bottom, making it less dense than the surrounding fluid, so that the blob floats to the surface, where it loses heat, becomes denser than the surrounding fluid and sinks to the bottom. The lava lamp model is not applicable to the Earth's mantle due to compression caused by over-burden weight (Herndon, 2009, 2010a, b).

Earth-mantle density as a function of radius is shown in Fig. 4. Because of the weight of the rock above, the mantle is about 62% more dense at the bottom than the top. The Rayleigh Number, as discussed by Herndon (2009), was derived on the basis of constant density and is thus inappropriate to apply to the Earth's compressed mantle.

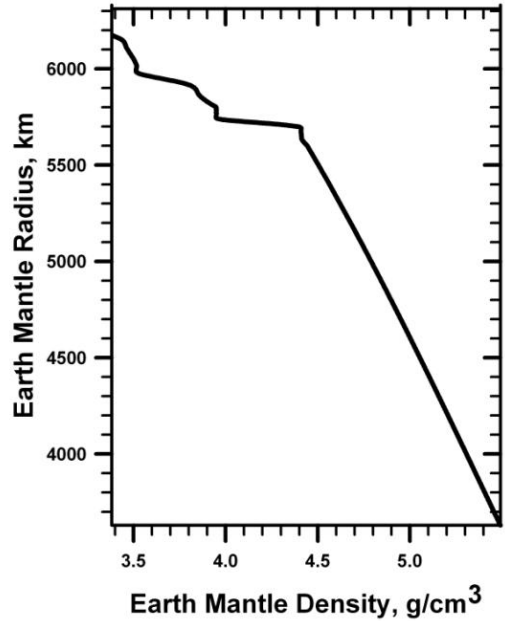

**Fig. 4** Mantle density as a function of Earth radius. Data from Dziewonski and Anderson (1981).

Consider a "parcel" of matter at the base of the Earth's lower mantle existing at the prevailing temperature, $T_0$, and having density, $\rho_0$, indicated by the data upon which Fig. 4 is based (Dziewonski and Anderson, 1981). Now, suppose that the "parcel" of bottom-mantle matter is selectively heated to temperature $\Delta T$ degrees above $T_0$. The "parcel" will expand to a new density, $\rho_z$, given by

$$\rho_z = \rho_0 (1-\alpha\Delta T)$$

where $\alpha$ is the volume coefficient of thermal expansion at the prevailing temperature and pressure.

Now, consider the resulting dynamics of the newly expanded "parcel". Under the assumption of ideal, optimum conditions, the "parcel" will suffer no heat loss and will encounter no resistance as it floats upward to come to rest at its "neutral buoyancy", the point at which its own density is the same as the prevailing mantle density. The Earth-radius of the "neutral buoyancy" point thus determined can be obtained from the data upon which Fig. 4 is based; the "maximum float distance" simply is the difference between that value and the Earth-radius at the bottom of the lower mantle.

The relationship between "maximum float distance" and $\Delta T$ thus calculated for the lower mantle, shown in Fig. 5, proves conclusively the physical impossibility of lower mantle convection. At the highest $\Delta T$ shown, the "maximum float distance" to the point of "neutral buoyancy" is <25 km, just a tiny portion of the 2900 km distance required for lower mantle convection, and the 3750 km required for whole-mantle convection. Even with the assumed "ideal, optimum conditions" and an unrealistically great ΔT = 600°K, an error in the value of α by two orders of magnitude would still not cause the "maximum float distance" to reach 2900 km, the top of the lower mantle.

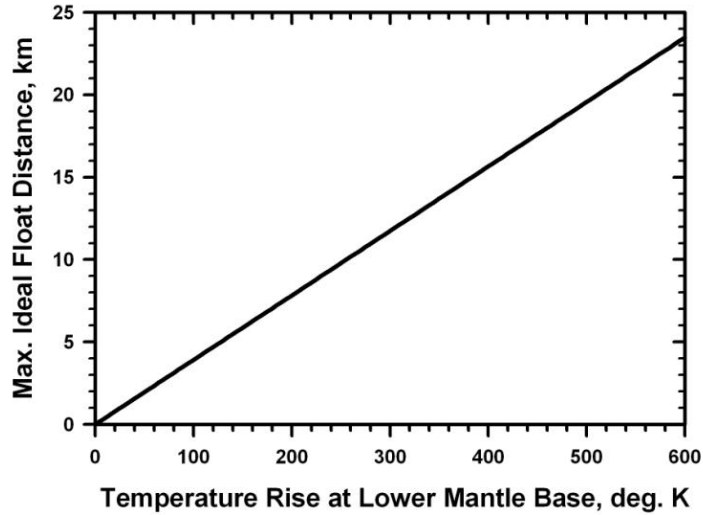

**Fig. 5** The "maximum float distance" to "neutral buoyancy" from the base of the lower mantle as a function of "parcel" temperature rise. The value used for the coefficient of thermal expansion, $\alpha=0.37\times10^{-5}$ K$^{-1}$, is from the standard reference state value of MgSiO$_3$ perovskite (Oganov et al., 2001), reduced by 80% to take into account lower mantle base temperature and pressure, according to (Birch, 1952).

The same reasoning can be applied to prove the physical impossibility of upper mantle convection. Here the upper mantle is taken to begin at the seismic discontinuity at radius 5700 km, which separates the endo-Earth (lower mantle plus core) from the matter above.

The relationship between "maximum float distance" and ΔT thus calculated for the upper mantle, shown in Fig. 6, proves conclusively that convection is physical impossibility in upper mantle. Even with the assumed "ideal, optimum conditions" and an unrealistically great ΔT = 600°K, an error in the value of α by an order of magnitude would still not cause the "maximum float distance" to reach the top of the upper mantle. Upper mantle convection, like lower mantle convection, is physically impossible.

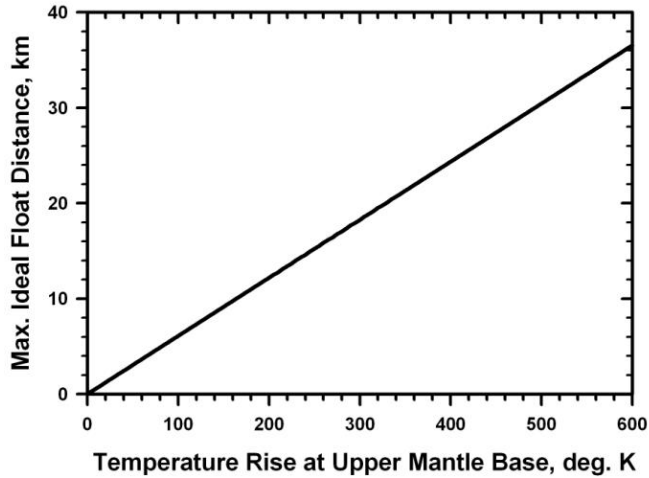

**Fig. 6** The "maximum float distance" to "neutral buoyancy" from the base of the upper mantle as a function of "parcel" temperature rise. The value used for the coefficient of thermal expansion, $\alpha=1.71\times10^{-5}$ K$^{-1}$, is for the γ phase of olivine (Meng et al., 1993).

The idea that a heated "parcel" of bottom mantle matter, under ideal, optimum circumstances will float upward to come to rest at its "neutral buoyancy", the point at which its own density is the same as the prevailing mantle density, is based upon the well-founded concept of submarine operation and design (Kormilitsin and Khalizev, 2001), which follows from Archimedes' discoveries. The abstraction is in assuming ideal, optimum conditions, i.e., the heated "parcel" will suffer no heat loss and will encounter no resistance as it floats upward. Because the mantle is solid, not liquid, the heated "parcel" would in fact suffer heat loss and encounter resistance; thus, the "maximum float distance" shown in Fig. 5 and in Fig. 6 would be severely reduced.

As proven above, mantle convection is physically impossible; thus, all investigations that depend upon or assume mantle convection are concomitantly refuted, which includes plate tectonics theory (Dietz, 1961;Hess, 1962;Le Pichon, 1968;Vine and Matthews, 1963), mantle convection models (Schubert et al., 2001), and models which assume mantle convection, such as heat emplacement at the base of the crust involving convection (Phillips and Coltice, 2010). Herndon has presented a fundamentally different development of geodynamics which does not require mantle convection (Herndon, 2010a, b).

# References

Birch, F.: Elasticity and constitution of the Earth's interior, J. Geophys. Res., 57, 227-286, 1952.

Bull, A. J.: A hypothesis of mountain building, Geol. Mag., 58, 364-397, 1921.

Dietz, R. S.: Continent and ocean basin evolution by spreading of the sea floor, Nature, 190, 854-857, 1961.

Dziewonski, A. M., and Anderson, D. A.: Preliminary reference Earth model, Phys. Earth Planet. Inter., 25, 297-356, 1981.

Herndon, J. M.: Uniqueness of Herndon's georeactor: Energy source and production mechanism for Earth's magnetic field, arXiv.org/abs/0901.4509, 2009.

Herndon, J. M.: Impact of recent discoveries on petroleum and natural gas exploration: Emphasis on India, Current Science, 98, 772-779, 2010a.

Herndon, J. M.: Inseparability of science history and discovery, Hist.Geo. Space Sci., 1, 25-41, 2010b.

Hess, H. H.: History of Ocean Basins, in: Petrologic Studies: A Volume in Honor of A. F. Buddington, Geological Society of America, Boulder, 599-620, 1962.

Holmes, A.: Radioactivity and Earth movements, Trans. geol. Soc. Glasgow 1928-1929, 18, 559-606, 1931.

Kormilitsin, Y. N., and Khalizev, O. A.: Theory of Submarine Design, Riviera Maritime Media, Enfield, UK, 339 pp., 2001.

Le Pichon, X.: Sea-floor spreading and continental drift, J. Geophys. Res., 73, 3661-3697, 1968.

Meng, Y., Weidner, D. J., Gwanmesia, D. G., Liebermann, R. C., Vaughan, T., Wang, Y., Leinenweber, K., Pacalo, R. E., Yeganeh-Haeri, A., and Zhao, Y.: In situ high P-T X ray diffraction studies of three polymorphs of $Mg_2SiO_4$, J. Geophys. Res., 98, 22199-22207, 1993.

Oganov, A. R., Brodholt, J. P., and Price, G. D.: Ab initio elasticity and thermal equation of state of $MgSiO_3$ perovskite Earth Planet. Sci. Lett., 184, 555-560, 2001.

Phillips, B. R., and Coltice, N.: Temperature beneath continents as a function of continental cover and convective wavelength, J. Geophys. Res., 115, B04408-B04421, 2010.

Schubert, G., Turcotte, D. L., and Olsen, P.: Mantle Convection in the Earth and Planets, Cambridge University Press, Cambridge, 940 pp., 2001.

Smith, D. G.: Display Devices, U. S. Patent 3,387,396, 3, 1968.

Snider-Pellegrini, A.: La Création et ses mystères dévoilés (Creation and its Mysteries Unveiled) Paris, 1858.

Vine, F. J., and Matthews, D. H.: Magnetic anomalies over oceanic ridges, Nature, 199, 947-949, 1963.

Wegener, A. L.: Die Entstehung der Kontinente, Geol. Rundschau, 3, 276-292, 1912.